\documentclass[prl,twocolumn,showpacs,amsmath,amssymb]{revtex4-1}

\usepackage{graphicx}
\usepackage{amsfonts}
\usepackage{amsmath}
\usepackage{amssymb}
\usepackage{color}
\usepackage{bm}
\usepackage{bbm}
\usepackage{enumerate}
\usepackage{amsfonts, amsmath, amsthm, amssymb} 
\usepackage{mathtools}
\usepackage{array}
\usepackage[colorlinks,bookmarks=false,citecolor=blue,linkcolor=red,urlcolor=blue]{hyperref}
\usepackage[usenames,dvipsnames,svgnames,table]{xcolor}
\newcommand{\be}{\begin{equation}}
\newcommand{\bee}{\begin{equation*}}
\newcommand{\ee}{\end{equation}}
\newcommand{\eee}{\end{equation*}}
\newcommand{\bearre}{\begin{eqnarray*}}
\newcommand{\eearre}{\end{eqnarray*}}
\newcommand{\bearr}{\begin{eqnarray}}
\newcommand{\eearr}{\end{eqnarray}}

\begin{document}

\title{ Reconstructing the quantum critical fan of strongly correlated systems via quantum correlations
}

\author{Ir\'en\'ee Fr\'erot$^{1,2}$\footnote{Electronic address: \texttt{irenee.frerot@icfo.eu}}, 
and Tommaso Roscilde$^{2,3}$ \footnote{Electronic address: \texttt{tommaso.roscilde@ens-lyon.fr}}
 }

\affiliation{$^1$ICFO-Institut de Ciencies Fotoniques, The Barcelona Institute of Science and Technology, Av. Carl Friedrich Gauss 3, 08860 Castelldefels (Barcelona), Spain}
\affiliation{$^2$ Univ Lyon, Ens de Lyon, Univ Claude Bernard, CNRS, Laboratoire de Physique, F-69342 Lyon, France}
\affiliation{$^3$ Institut Universitaire de France, 103 boulevard Saint-Michel, 75005 Paris, France}
\date{\today}

%----------------------------------------------------------------------------------------
%	ABSTRACT
%----------------------------------------------------------------------------------------

\begin{abstract}
Albeit occurring at zero temperature, quantum critical phenomena are known to have a huge impact on the finite-temperature phase diagram of strongly correlated systems -- an aspect which gives experimental access to their observation.  In particular the existence of a gapless, zero-temperature quantum critical point is known theoretically to induce the existence of an extended region in parameter space -- the so-called quantum critical fan -- characterized by power-law temperature dependences of all observables, with exponents related to those of the quantum critical point. Identifying experimentally the quantum critical fan and its crossovers to the other regions (renormalized classical, quantum disordered) remains nonetheless a big challenge. Focusing on paradigmatic models of quantum phase transitions, here we show that quantum correlations - captured by the quantum variance of the order parameter (I. Fr\'erot and T. Roscilde, Phys. Rev. B {\bf 94}, 075121 (2016)) - exhibit the temperature scaling associated with the quantum critical regime over an extended parameter region, much broader than that revealed by ordinary correlations, and with well-defined crossovers to the other regimes. The link existing between the quantum variance and the dynamical order-parameter susceptibility paves the way to an experimental reconstruction of the quantum critical fan using \emph{e.g.} spectroscopy on strongly correlated quantum matter.  
\end{abstract}
 \maketitle
 
\begin{center}
 \begin{figure}[ht!]
 \includegraphics[width=0.6\columnwidth]{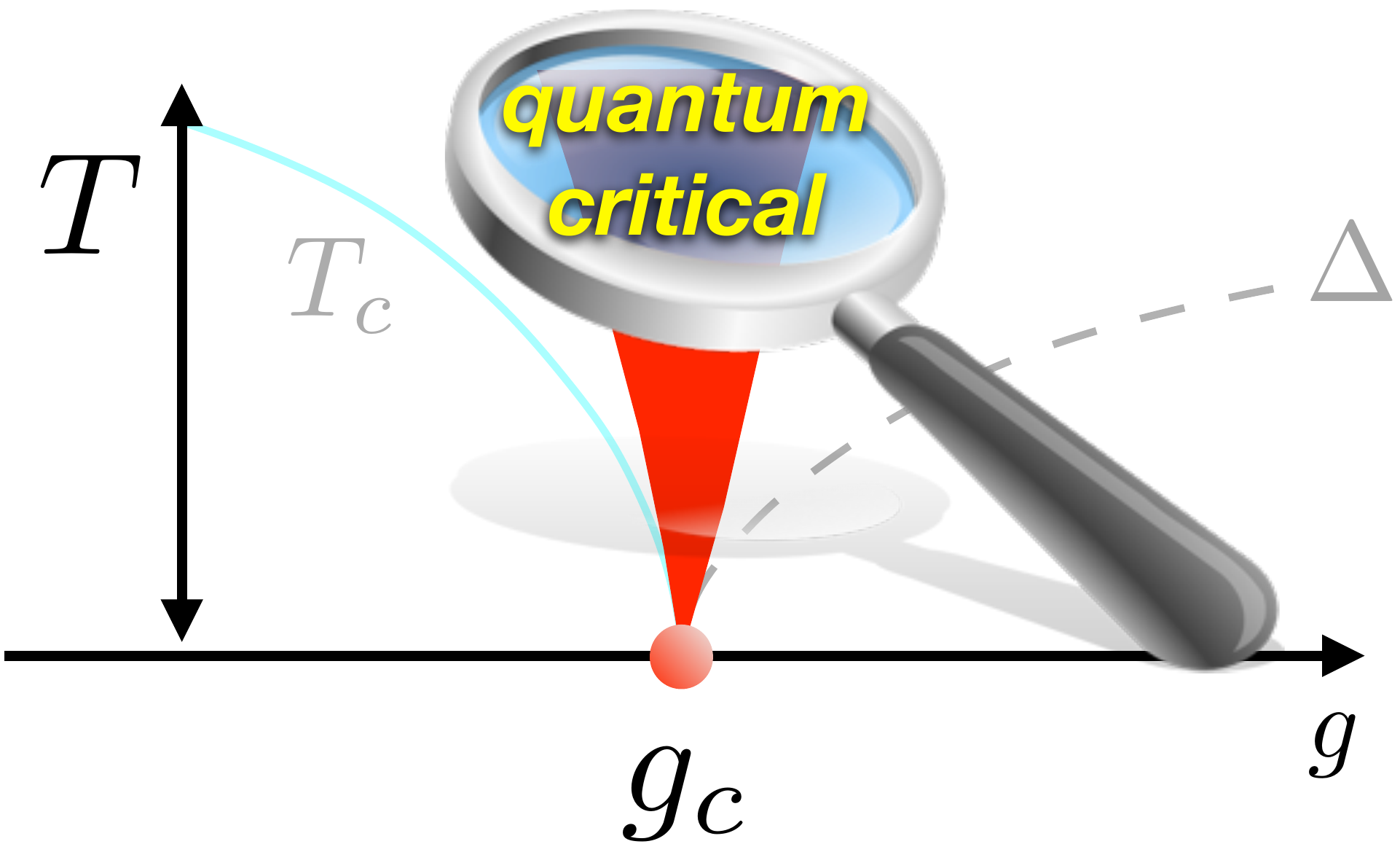}
 \caption{The quantum critical fan can be seen effectively as a "magnifying lens" for the quantum critical point, making its existence observable over an extended range of temperatures and of the control parameter $g$ of the transition.}
 \label{f.lens}
 \end{figure}
 \end{center}

 \emph{Introduction.} Quantum critical phenomena \cite{Sondhietal1999,Sachdevbook,Continentinobook,Duttabook,Carrbook} represent possibly the most dramatic manifestation of quantum mechanics at the macroscopic scale. Their typical setting involves an Hamiltonian ${\cal H}  = {\cal H}_0 + g V$ in which the competition between the two non-commuting terms ${\cal H}_0$ and $V$, controlled by the parameter $g$, induces a macroscopic rearrangement of the ground state at a critical value $g_c$, accompanied by the appearance of critical quantum fluctuations of collective observables at all length scales. 
 %The divergence of the range of ground-state correlations, controlled by an Hamiltonian parameter, is naturally accompanied by a strong enhancement of entanglement among the real-space degrees of freedom, as captured \emph{e.g.} by singularities in the von Neumann entanglement entropy of a spatial bipartition \cite{Franchinietal2007,Metlitskietal2009,FrerotR2016-2}. 
 This complex behavior of correlation and entanglement properties emerges from extensive theoretical work based on exactly solvable microscopic models \cite{Duttabook}, quantum field theory \cite{Sachdevbook} as well as numerical studies \cite{Kauletal2014}. Experiments generally do not have access to ground-state physics, but it has soon been realized \cite{CHN1989} that zero-$T$ quantum critical points affect a sizable portion of the finite-$T$ phase diagram, by inducing the presence of a so-called \emph{quantum critical (QC) regime}, whose thermodynamics is completely controlled by the quantum critical point (QCP). Indeed observables in the QC regime are expected to exhibit a power-law dependence on temperature with exponents descending from the critical exponents at the QCP (hereafter referred to as thermal QC scaling). Strikingly, the QC regime is expected to be \emph{wider} in parameter space, the higher the temperature: namely, as sketched in Fig.~\ref{f.lens}, it acts as a ``magnifying lens'' for the QCP. Even more strikingly, the finite-$T$ QC regime ignores completely the physics of the $T=0$ and low-$T$ phases at $g\neq g_c$ \cite{Chubukovetal1994}, which are generally a) an ordered phase with a classical analog (for, say, $g<g_c$); and b) a gapped quantum disordered phase (for $g>g_c$). This implies that, if the temperature is lowered from a point at $g\neq g_c$ in the QC regime, a crossover must occur towards a thermodynamic regime which is instead controlled by the presence of long-range order in the ground state -- the so-called renormalized classical (RC) regime for $g<g_c$ -- or by the presence of a gap above a disordered ground state (the QD regime for $g>g_c$). This is all the more striking, as it shows that a strictly quantum $T=0$ phenomenon (the QCP), governed by divergent quantum fluctuations, can have consequences on the phase diagram at temperatures $T$ which are higher than those necessary to melt long-range order via a classical thermal transition. 
 Many exciting platforms for the exploration of quantum critical phenomena can be found across the physical spectrum \cite{Carrbook,Gegenwartetal2008,Trotzkyetal2010,Endresetal2012,Zhangetal2012,Duttabook}. But can one reconstruct the QC regime quantitatively? The special scaling properties of the thermodynamics and the dynamical response functions at $g=g_c$ and finite $T$ (along the so-called QC trajectory) have been observed in several systems, including magnetic insulators  \cite{Lakeetal2005,Mukhopadhyayetal2012,Kinrossetal2014,Povarovetal2015,halgetal2015} and heavy-fermion compounds \cite{Schroederetal2000,Gegenwartetal2008}; but the persistence of the QC regime away from $g_c$, and its crossover into the competing low-$T$ regimes, are almost uniquely observed via transport properties in heavy-fermion materials \cite{Gegenwartetal2008} -- the ``strange metal" phase in cuprate superconductors is also interpreted as an extended QC regime \cite{Sachdev2010} associated with a putative QCP \cite{Ramshawetal2015,Badouxetal2016}. Hence it is fair to say that the quantitative extent of the QC regime, and its crossovers towards the RC and QD regime, remain challenging to observe. Quite remarkably, the same observations can be repeated for theoretical calculations on microscopic models, for which the quantitative extent of the QC regime is rarely investigated \cite{KoppC2005}. A general scenario (corroborated by the present work) is that different observables exhibit thermal QC scaling over different regions in the $(g,T)$ parameter space. Therefore it is crucial to identify those observables which manifest such a scaling over the broadest possible range.  
 %It is tempting to attribute the difficulties in identifying the QC regime to the rather qualitative character of its very definition \cite{Chubukovetal1994,Sachdevbook,SachdevK2010}. 

  \begin{center} 
\begin{figure} [ht!]
\includegraphics[width=0.8\columnwidth]{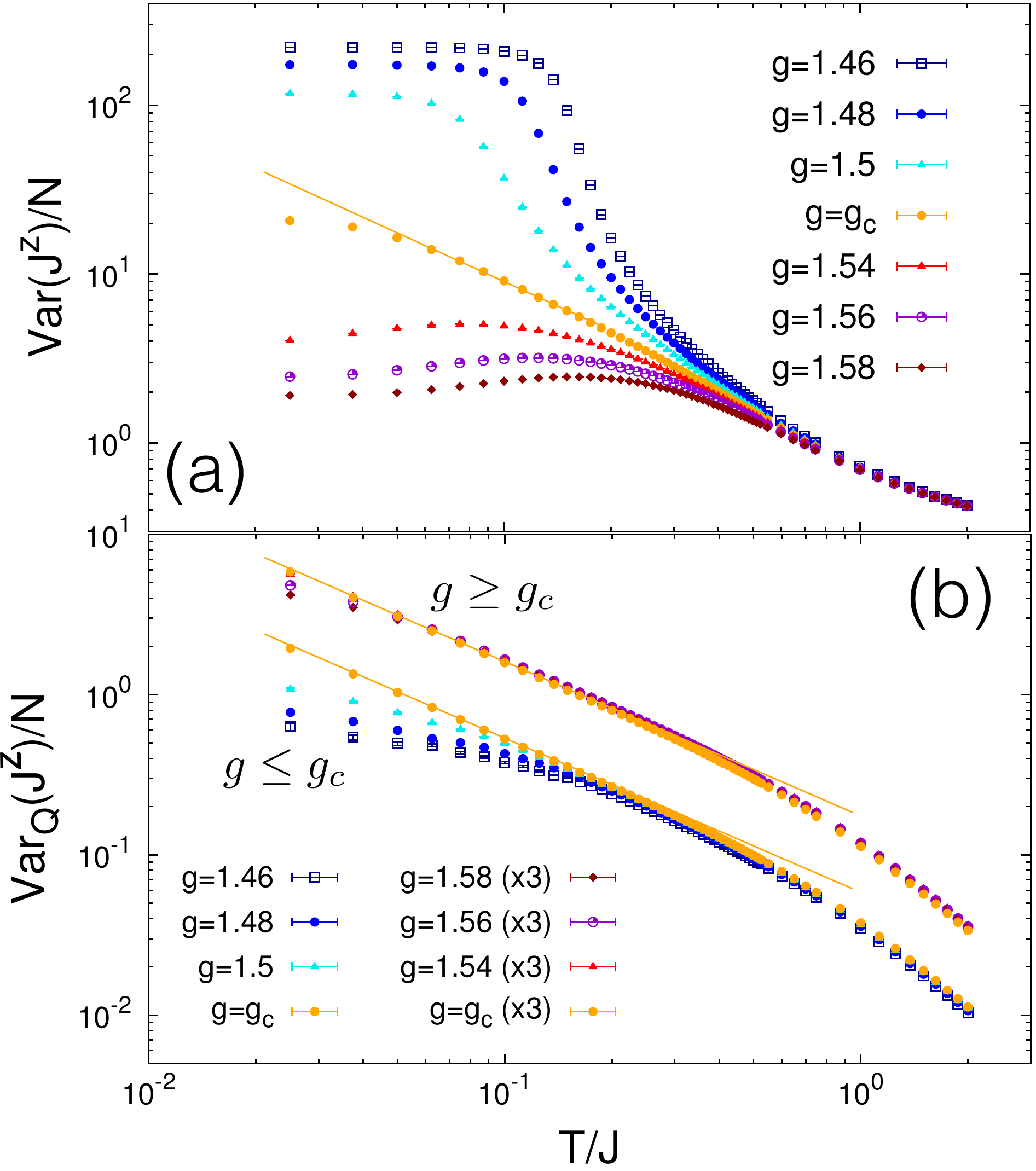}
\caption{Temperature scaling of the total variance (a) and of the quantum variance (b) of the order parameter around the quantum critical point of the 2d quantum Ising model (data have been obtained on a $L=64$ lattice). The quantum variance above the quantum critical point has been multiplied by a factor of 3 to improve readability. Solid lines are the QC scaling forms $G(0)T^{-\psi}$ (in (a)) and $G^{(Q)}(0)T^{-\psi}$ (in (b)) respectively.} 
\label{f.Tscaling-Ising}
\end{figure} 
\end{center} 
  
  \emph{QC regime from quantum correlations.} Here we propose a constructive definition of the QC regime based on observables that do not admit any classical analog, namely quantum coherence measures, capturing quantum correlations and fluctuations for generic mixed states. At first sight this sounds very logical: the $T=0$ QCP is characterized by critical quantum fluctuations, and the QC regime, if regarded as ``echoing" the QCP at finite $T$, should be characterized by enhanced quantum fluctuations as well. The quantum coherence measures of our interest generally belong to the family identified by Petz \cite{petz1996, streltsovetal2017} as generalizations of the quantum Fisher information (QFI) \cite{BraunsteinC1994, PezzeS2014}. Among this family of quantities, we focus on the recently proposed \emph{quantum variance} (QV) \cite{FrerotR2016} of an observable $O$, possessing a simple definition at thermal equilibrium at inverse temperature $\beta = (k_B T)^{-1}$; it is the difference between the (total) variance (TV) ${\rm Var}[O] = \langle O^2 \rangle - \langle O \rangle^2$ and the susceptibility
  \begin{equation}
  {\rm Var}_Q[O] =  {\rm Var}[O] - k_B T \chi_{O}
  \label{e.QV}
  \end{equation}  
  where  $\chi_{O} =   (\partial \langle O \rangle/\partial h)_{h=0} $ and $h$ is a field coupling to $O$ in the Hamiltonian as ${\cal H} - h O$.  Beyond its transparent physical meaning (difference between fluctuations and response function), the QV (like the QFI) has the fundamental property of being an entanglement witness -- denying separability of the state of the system into clusters of size $k$ (or smaller) when it exceeds a $k$-dependent bound \cite{FrerotR2016, Hyllusetal2012, Toth2012, Pezzeetal2016}. Moreover, unlike the QFI, it has the remarkable property of being directly accessible to state-of-the-art calculations for equilibrium quantum many-body systems at finite $T$, as \emph{e.g.} worldline quantum Monte Carlo \footnote{
  The computation of the QFI requires the precise knowledge of the dynamical response function at real frequencies \cite{Haukeetal2016}, which can only be inferred from quantum Monte-Carlo data through analytical continuation, an operation very sensitive to numerical noise.}. This elevates the QV to the observable of choice to explore quantum coherence properties across the phase diagram of quantum-critical phenomena.

  \begin{center} 
\begin{figure*} [ht!!]
\includegraphics[width=0.9\textwidth]{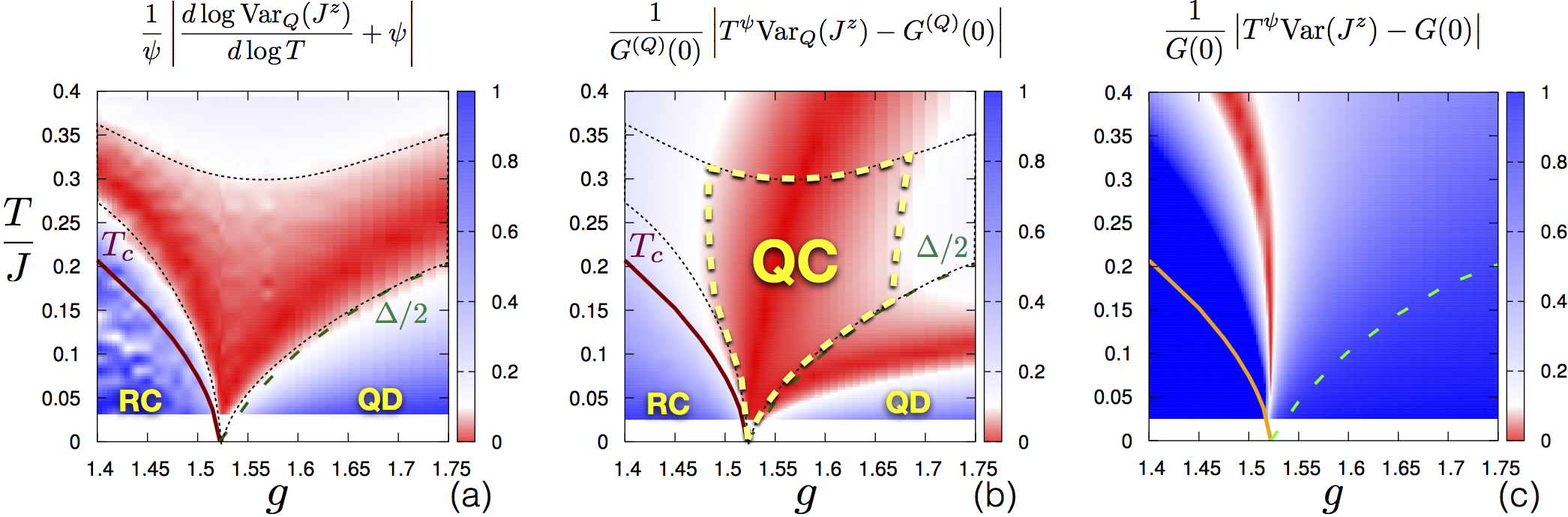}
\caption{\emph{Reconstruction of the quantum critical fan of the 2d TFI model via the quantum variance.} (a) Relative deviation of the logarithmic derivative of the order-parameter QV from the QC scaling exponent $\psi$ -- the data shown have been obtained for a $L=64$ lattice. The solid and dashed lines indicate the critical temperature \cite{HesselmannW2016} and half of the spectral gap (extracted from the $T$ scaling of $\langle S^x\rangle$) respectively, while the dotted line marks the region with less than 10\% deviation; (b) Relative deviation of ${\rm Var}_Q(J^z) T^\psi$ from the QC amplitude $G^{(Q)}(0)$; same symbols as in (a). The dashed yellow line encircles the region with less than 10\% deviation on both the prefactor and on the logarithmic derivative; (c) Relative deviation of ${\rm Var}(J^z) T^\psi$ from the QC prefactor $G^{(Q)}(0)$; other symbols as in (a).}
\label{f.QCfan-Ising}
\end{figure*} 
\end{center}    
  
  Both the variance and the susceptibility in Eq.~\eqref{e.QV} can be expressed as integrals of the imaginary part of the dynamical susceptibility \cite{Foersterbook}, resulting in the fundamental relationship:
  \begin{equation}
  {\rm Var}_Q[O] = \hbar \int_0^\infty \frac{d\omega}{\pi} {\cal L}(\beta \hbar \omega /2) ~\chi''_{O}(\omega)~.
  \label{e.chi2nd}
  \end{equation}
  where ${\cal L}(x) = \coth x - 1/x$ is the Langevin function. Since ${\cal L}(x) \to x/3$ for $x \to 0$, one sees that ${\rm Var}_Q[O]$ is dominated by modes with frequency $\omega$ such that $\beta\hbar\omega \gtrsim 1$, namely modes which are mildly (or not at all) affected by thermal fluctuations. A similar expression to Eq.~\eqref{e.chi2nd} holds for the QFI, with the replacement ${\cal L}(x) \to 4\tanh(x)$ \cite{Haukeetal2016}.
  When $O$ is the \emph{order parameter} of the quantum phase transition of interest, the dynamical susceptibility in the vicinity of the QCP is expected to obey the scaling form $\chi''_{O}(\omega) = T^{-(2-\eta)/z} \Phi_{O}  \left [  (g-g_c)^{\nu z}/T, \omega/T \right ]$ (where $\eta, \nu$ and $z$ are the correlation function, correlation length and dynamical critical exponent of the QCP respectively, and $\Phi_O$ is a universal function up to a prefactor)  \cite{Sachdevbook}. This directly translates into a scaling Ansatz for the QV:
  \begin{equation}
  {\rm Var}_Q[O] = T^{-\psi} G^{(Q)}_O[(g-g_c)^{\nu z}/T]
  \label{e.scalingQV}
  \end{equation}
  where $\psi = (2-\eta)/z-1$ and $G^{(Q)}_O \sim \int d\omega ~{\cal L}(\beta \hbar \omega /2) ~\Phi_O$. The TV of the order parameter ${\rm Var}[O]$ possesses a similar scaling form to Eq.~\eqref{e.scalingQV}, but with a different scaling function $G^{(Q)}_O \to G_O \sim  \int d\omega ~\coth(\beta \hbar \omega /2) ~\Phi_O$.  

 Eq.~\eqref{e.scalingQV} forms the basis of our constructive definition of the QC regime as detected by quantum correlations. Being controlled by the QCP alone, such a regime must be nearly insensitive to whether the control parameter $g$ lies above or below $g_c$. This defining condition requires that, in the QC regime, the scaling function $G^{(Q)}_O(x)$ depend very weakly on its argument, namely $G^{(Q)}_O(x) \approx G^{(Q)}_O(0)$. This leads us then to the following \emph{quantitative definition} for the QC regime in the $(g,T)$ plane: 
 \begin{equation}  
{\rm QC ~regime:} ~~~ {\rm Var}_Q[O](g,T) \approx T^{-\psi} G^{(Q)}_O(0)
\label{e.QC}
 \end{equation}
 where the $\approx$ sign implies that the above condition is satisfied within some tolerance. As the QC regime is \emph{not} a phase of matter which is divided from competing phases by sharp boundaries, the tolerance defines operatively the crossover lines towards the other regimes (RC and QD) in the vicinity of the QCP. The constructive definition of the QC regime offered by Eq.~\eqref{e.QC} identifies the latter regime with the region in the $g$-$T$ phase diagram in which the \emph{$T$-dependence of the quantum fluctuations of the order parameter is uniquely controlled by the presence of the QCP} -- namely it is the same (up to some tolerance) as along the QC trajectory (variable $T$ at $g=g_c$). The very fact that such a regime exists in an extended, fan-shaped region, is a fundamental test of the validity of our definition.  A similar condition could obviously be formulated for the more conventional TV in the form ${\rm Var}[O] \approx T^{1- \frac{2-\eta}{z}} G_O(0)$: as we shall see shortly, this condition in practice singles out only the QC trajectory. 

\emph{2$d$ transverse field Ising model.}  We demonstrate our constructive definition of the QC regime using two paradigmatic examples of quantum phase transitions in quantum spin models. We begin by considering the 2$d$ transverse field Ising (TFI) model \cite{Duttabook}
  \begin{equation}
  {\cal H}/J = -\sum_{\langle ij \rangle} S_i^z S_j^z - g \sum_i S_i^x  
  \end{equation}  
  where the indices $\langle ij \rangle$ and $i$ run over the nearest-neighbor bonds and the sites of a square lattice, respectively; and $S_i^{\alpha}$ ($\alpha = x,y,z$) are $S=1/2$
spin operators. A critical value $g_c=1.522...$ \cite{BloeteD2002} of the transverse field divides a ferromagnetic regime ($g < g_c$) from a quantum paramagnetic one ($g>g_c$). We calculate the equilibrium properties of this model on $N=L\times L$ lattices with periodic boundary conditions numerically using Stochastic Series Expansion quantum Monte Carlo \cite{Sandvik2010,FrerotR2017}, which gives direct access to the TV and QV of most relevant observables \cite{FrerotR2016}. The temperature scaling of the variances (total and quantum) of the macroscopic order parameter $J^z = \sum_i S_i^z$ in the vicinity of the QCP are contrasted in 
Fig.~\ref{f.Tscaling-Ising}.  Here we take for simplicity ${\rm Var}(J^z) =  \langle (J^z)^2 \rangle$ as $\langle J^z \rangle =0$ on finite lattices. We have checked that our conclusions do not change when considering a finite-size estimate of the actual variance, namely $\langle (J^z)^2 \rangle - N^2 m_L^2$ where $m_L^2 = \langle S_i^z S_{i+L/2}^z \rangle$.
The QCP appears to control the $T$ dependence of ${\rm Var}(J^z)$ only along the QC trajectory [Fig. \ref{f.Tscaling-Ising}(a)], where it exhibits the expected power-law dependence $\sim T^{-\psi}$; on the contrary the TV is strongly bent upward by the finite-$T$ transition for $g < g_c$, as well as downward by the opening of a spin gap for $g > g_c$. 

 This picture is completely changed when one looks at quantum fluctuations. Indeed Fig.~\ref{f.Tscaling-Ising}(b) shows that a power-law QC scaling of the QV as $\sim T^{-\psi}$ is manifested not only along the QC trajectory (down to $T=0$), but it can be observed also over sizable segments of the $T$-dependence both above and below the QCP. For $g< g_c$ this is due to the fundamental property of the QV to be nearly insensitive to finite-temperature transitions \cite{FrerotR2016} -- only \emph{weak} singularities, in the form of inflection points, can appear at $T_c$ \cite{Frerotetal2018}. Therefore the so-called Ginzburg region \cite{Sachdevbook} (in which thermal criticality dominates the behavior of the system) is minute in the $T$-dependence of the QV. Interestingly, a similar observation also applies to the case $g > g_c$: unlike thermal fluctuations, quantum fluctuations are much more moderately suppressed by the opening of a gap. This observation can be understood by considering that, associated to the quantum fluctuations of the order parameter, there is an intrinsic quantum coherence length $\xi_Q$ \cite{MalpettiR2016} which is always finite at finite $T$, and much smaller than the ordinary correlation length $\xi$. Approaching the QCP, $\xi \approx c T^{-1/z} = c/T$, but the opening of a gap ($\Delta$) for $g \neq g_c$ cuts off the quantum-critical growth, as $\xi$ saturates to its ground-state value $\xi(T=0) \approx c/\Delta$. Such saturation occurs for $T \approx \Delta$. Similarly, one expects that  $\xi_Q \approx c_Q/T$, but with $c_Q \ll c$; nonetheless, at $T=0$, $\xi_Q$ saturates to the same value $\xi_Q(T=0)=\xi(T=0)=c/\Delta$. Therefore, the saturation occurs at lower temperatures, $T \approx (c_Q / c) \Delta$. In the Supplementary Material (SM) \cite{SM}, we show that $c_Q/c \approx 1/6$. Hence, one may expect that saturation temperatures for $\xi$ and $\xi_Q$ are in a similar proportion in the gapped phase.    
 In summary, both above and below $g_c$ the QV exhibits a clear crossover from a power-law regime varying as $T^{-\psi}$ to a saturating regime, occurring around a temperature $T \sim T_c$ or $T\sim \Delta$: this behavior reveals then the crossover from the QC to the RC and QD regimes. Hence we can deduce that the QCP controls the thermodynamics of quantum fluctuations over a region of sizable width, and which is fan-shaped (namely broader in $g$, the higher $T$).   
  
   In order to quantitatively reconstruct this fan-shaped region in which the $T$-dependence of the QV is influenced by the existence of the QCP, we establish the following criteria: 1) the QV exhibits a power-law dependence in $T$ with an exponent (namely its logarithmic derivative) reproducing $\psi = 0.964...$ \cite{PelissettoV2002} (within some tolerance $\epsilon$); 2) the coefficient of the power-law dependence (estimated as ${\rm Var}_Q(J^z) T^{\psi}$) reproduces the one along the QC trajectory $G^{(Q)}(0)$  (within the same tolerance $\epsilon$). These two criteria are illustrated in Fig.~\ref{f.QCfan-Ising}(a)-(b):  the region matching both criteria is then identified as the QC regime of quantum fluctuations. Obviously the extent of such a regime in the phase diagram depends crucially on $\epsilon$ (taken as 10\% in Fig.~\ref{f.QCfan-Ising}): yet it is important to observe that, regardless of the value of $\epsilon$, its lower boundaries, marking the onset of the crossovers towards the RC and QD regimes, follow faithfully the temperature scales set by $T_c$ and $\Delta$ -- both scaling as $|g-g_c|^{\nu z}$ in the vicinity of the QCP. In contrast, applying similar criteria to the scaling of ${\rm Var}(J^z)$ essentially reconstructs the QC trajectory only, as already anticipated above.  In the SM \cite{SM} we provide evidence of a similar phenomenology of the QV, as well as of the QFI, for the exactly solvable cases of the TFI in $d=1$ and $d=\infty$.  
    
\begin{center} 
\begin{figure} [ht!]
\includegraphics[width=0.7\columnwidth]{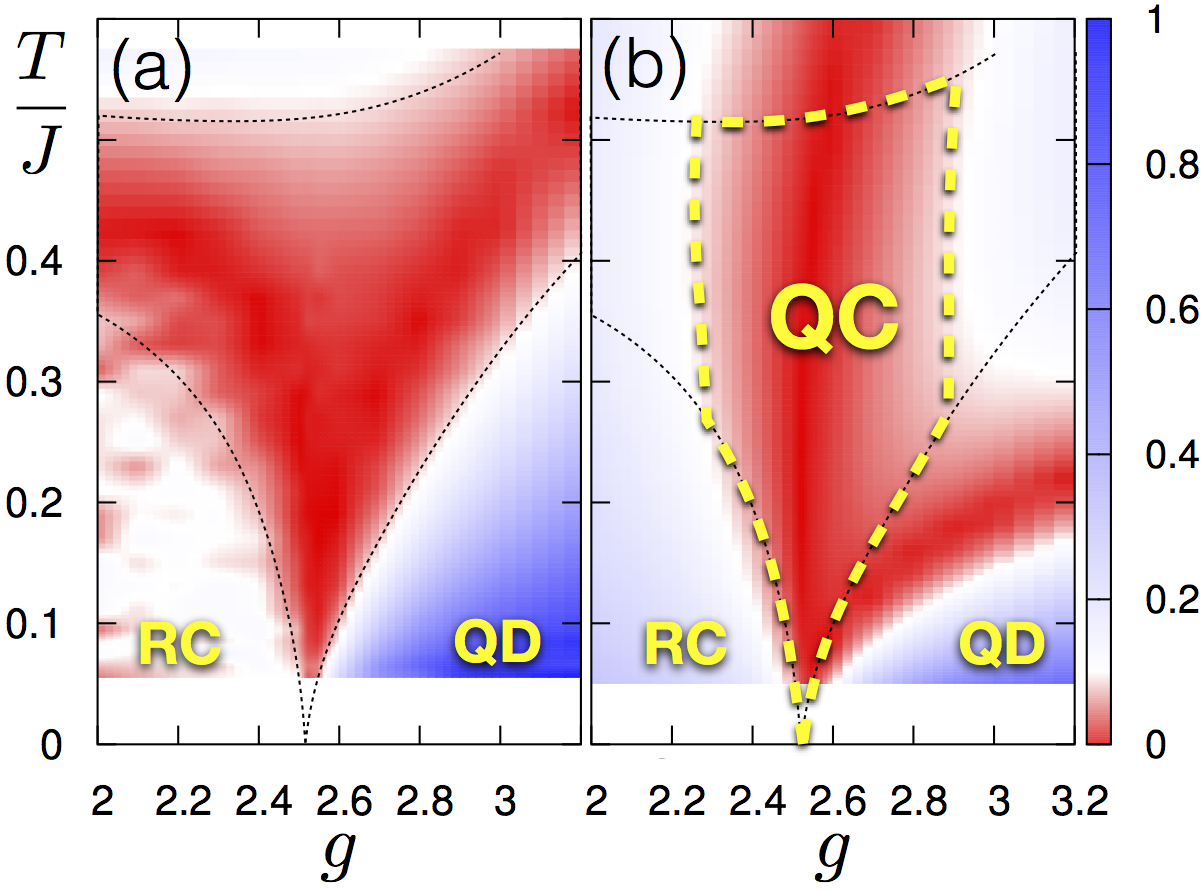}
\caption{\emph{Reconstruction of the QC fan of the Heisenberg bilayer.} Same significance of symbols as in panels (a) and (b) of Fig.~\ref{f.QCfan-Ising}.}  
\label{f.QCfan-Heis}
\end{figure} 
\end{center}   
  
\emph{Heisenberg bilayer.} We conclude by considering another paradigmatic model of quantum phase transitions, namely the $S=1/2$ Heisenberg antiferromagnet on a bilayer, with Hamiltonian ${\cal H}/J = \sum_{\langle ij \rangle} {\bm S}_i \cdot {\bm S}_j + g \sum_{\langle lm \rangle_{\perp}} {\bm S}_l \cdot {\bm S}_m$ comprising intralayer ($\langle ij \rangle$) as well as interlayer ($\langle lm \rangle_{\perp}$) bonds. A quantum phase transition at $g_c=2.522...$ \cite{SandvikS1994} divides a N\'eel antiferromagnetic regime (with order parameter given by the staggered magnetization $J^z_{\rm st} = \sum_i (-1)^i S_i^z$) from a non-magnetic dimer-singlet regime. The same kind of analysis of the order-parameter QV, as the one presented above for the 2$d$ TFI model, leads to Fig.~\ref{f.QCfan-Heis}. There the QC region, where the thermal behavior of quantum fluctuations is governed by the quantum critical point, is shown to be very broad (with a maximum width $\Delta g$ which is around 20\% of $g_c$) -- on the contrary, a similar analysis based on the behavior of the TV (not shown) singles out only a narrow region around the QC trajectory. In the case at hand the QC-RC crossover is marked by a crossover in the QV from the $T^{-\psi}$ power-law behavior into another power-law behavior, as the RC regime is gapless and without a finite-$T$ phase transition; while the QC-QD crossover is similar to the one observed in the 2$d$ TFI model (low-$T$ saturation of the QV). Despite its sizable width, the QC regime of the QV remains limited to a finite $g$ range and it does not come close to the limit $g=0$ corresponding to the most investigated limit of the 2$d$ Heisenberg antiferromagnet (2$d$HAF) -- this holds true even when restricting uniquely to the criterion of the logarithmic derivative. The 2$d$HAF Heisenberg model has been the subject of an intense search for signatures of a QC-RC crossover in the past \cite{Chubukovetal1994,Imaietal1993,Grevenetal1995,Roennowetal1999,KimT1998}: our data \cite{SM} exclude that such a crossover is visible in the QV of the order parameter.         
  
 \emph{Conclusions.} We have demonstrated that the existence of a zero-temperature quantum critical point (QCP) fully controls the thermodynamics of the quantum fluctuations of the order parameter (estimated via the quantum variance) in a broad, fan-shaped region above the QCP itself. Such a region can be identified with the elusive quantum critical (QC) regime, acting as a finite-$T$ magnifying lens of zero-$T$ quantum criticality. The extent of the QC regime, as revealed by quantum fluctuations, far exceeds that of conventional fluctuations properties - the latter contain a large thermal component subject to thermal criticality on one side of the QCP, and to a large suppression due to the opening of a gap on the other side. Therefore we open the unconventional perspective of using a property which bears \emph{no classical analog} to unveil a \emph{finite-temperature} regime - somewhat reminiscent of the use of entanglement to characterize QCPs at $T=0$.  Our findings are immediately applicable to detect the QC regime in numerical as well as field-theoretical studies of QC phenomena that have naturally access to the quantum variance. Most importantly, the quantitative relationship between the dynamical susceptibility and quantum fluctuations offers the possibility to access the latter in spectroscopic experiments on strongly correlated materials (neutron spectroscopy, AC susceptometry, etc.). This can serve as an effective tool to unveil the existence of zero-$T$ QCPs via finite-$T$ experiments, especially in situations in which the direct observation of the QCP proves elusive \cite{Sachdev2010,Shibauchietal2014}.
   
\emph{Acknowledgements.} We thank L. Pezz\'e and A. Smerzi for fruitful discussions, and for informing us of their recent preprint \cite{Gabbriellietal2018} on related work about the quantum Fisher information around quantum critical points. This work was supported by ANR ('ArtiQ' project). Numerical simulations were run on the PSMN cluster (ENS de Lyon). I. F. acknowledges support from the Spanish Ministry MINECO (National Plan 15 Grant: FISICATEAMO No. FIS2016-79508-P, SEVERO OCHOA No. SEV-2015-0522), Fundaci\'o Cellex, Generalitat de Catalunya (AGAUR Grant No. 2017 SGR 1341 and CERCA/Program), ERC AdG OSYRIS, EU FETPRO QUIC, and the National Science Centre, Poland-Symfonia Grant No. 2016/20/W/ST4/00314.  
       
 \newpage

\begin{center}
{\bf SUPPLEMENTARY MATERIAL}

{\bf \emph{Reconstructing the quantum critical fan of strongly correlated systems via quantum correlations}}

\end{center}

\section{Quantum coherence length vs. correlation length around the QCP}

\begin{center} 
\begin{figure} [ht!]
\includegraphics[width=0.8\columnwidth]{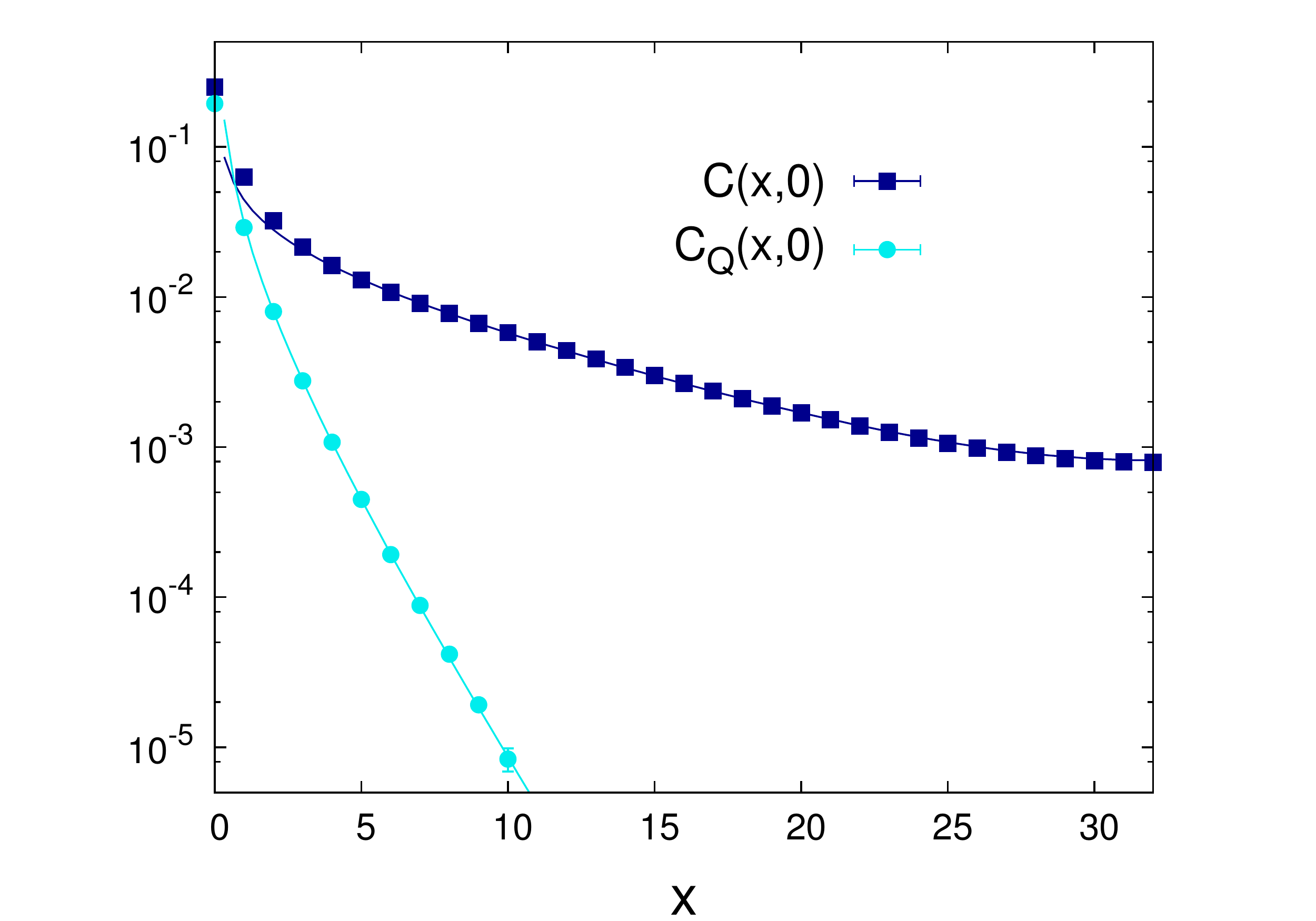}
\caption{Ordinary vs. quantum correlation functions for the 2$d$ transverse-field Ising model. The model parameters are $g = g_c$, $T/J = 0.1$ and $L=64$. The solid lines show fits to the form Eq.~\eqref{e.fit}.}  
\label{f.CCQ}
\end{figure} 
\end{center}

In this section we discuss the scaling of the correlation length and of the quantum coherence length in the QC regime of the 2$d$ transverse field Ising model. 
We extract the correlation length $\xi$ and the quantum coherence length $\xi_Q$ by fitting the ordinary correlation and the quantum correlation function \cite{MalpettiR2016}: 
\begin{eqnarray}
C({\bm r}) & = & \langle S_i^z S_{i+{\bm r}}^z \rangle \nonumber \\
C_Q({\bm r}) & = & \langle S_i^z S_{i+{\bm r}}^z \rangle - \frac{1}{\beta} \int_0^{\beta} d\tau \langle S_i^z(\tau) S_{i+r}^z(0) \rangle~.
\end{eqnarray} 
where $S_i^z(\tau) = e^{\tau \cal H} S_i^z e^{-\tau \cal H}$ is the imaginary-time evolved operator. For separations ${\bm r} = (x,0)$ both functions can be fitted to the form
\begin{equation}
f(x) = a \left ( \frac{e^{-x/\xi_{(Q)}}}{x^b} + \frac{e^{-(L-x)/\xi_{(Q)}}}{(L-x)^b} \right )~.
\label{e.fit}
\end{equation} 
An example of the fits are shown in Fig.~\ref{f.CCQ}, highlighting as well the vast difference in range between ordinary and quantum correlations. 

 \begin{center} 
\begin{figure} [ht!]
\includegraphics[width=0.8\columnwidth]{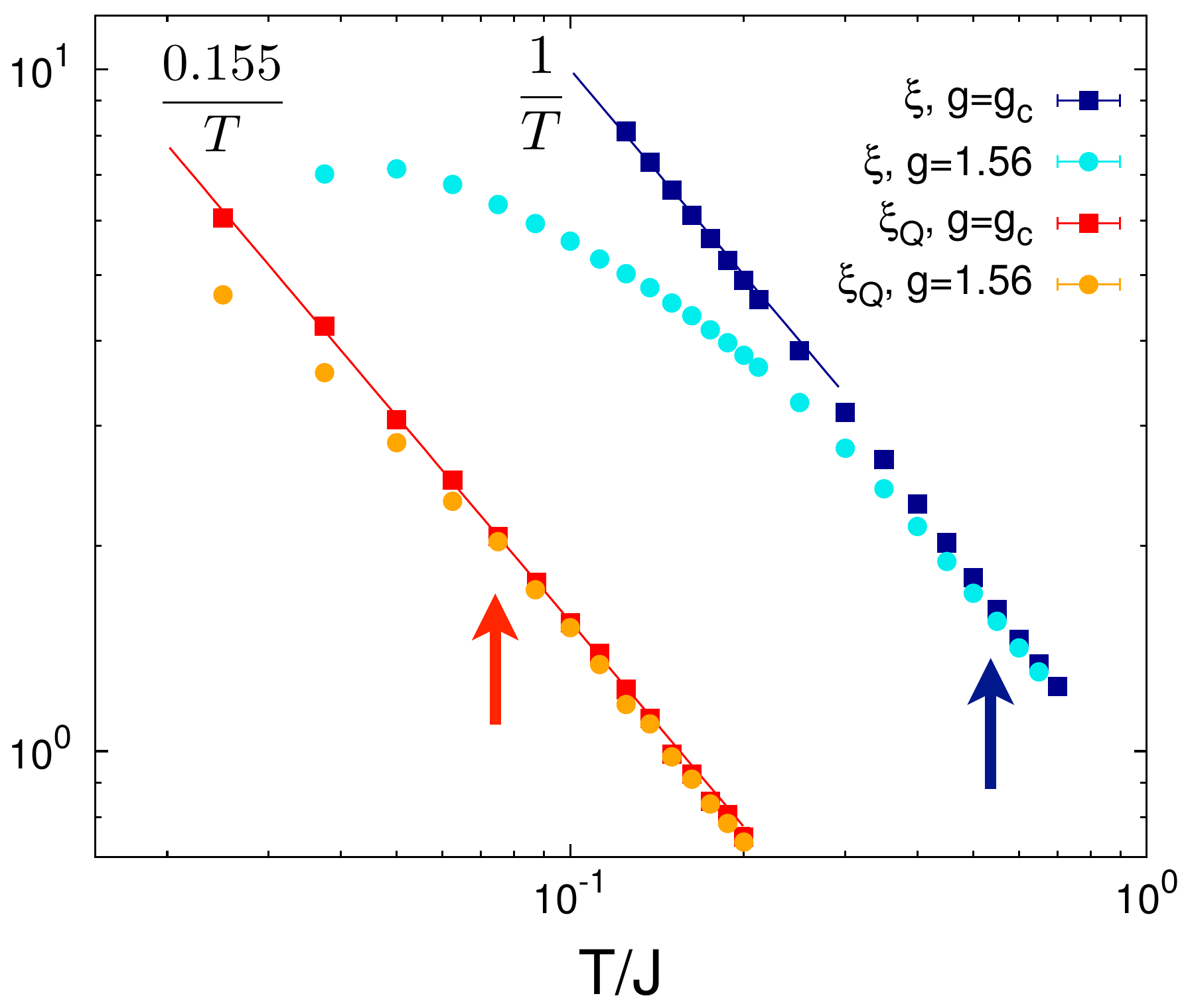}
\caption{Temperature scaling of the correlation length ($\xi$) and quantum coherence length ($\xi_Q$) along the quantum critical trajectory ($g=g_c=1.522...$) of the 2d transverse-field Ising model, and slightly above ($g=1.56$). The solid lines show the low-$T$ behavior as $\sim 1/T^{1/z}$ along the quantum critical trajectory -- notice the wide separation between $\xi$ and $\xi_Q$. The arrows show the temperatures around which $\xi$ (blue arrow) and $\xi_Q$ (red arrow) at $g=1.56$ depart from the same curves for $g=g_c$.}  
\label{f.xiTscaling-Ising}
\end{figure} 
\end{center}

The results of the fit of the above correlation functions along the quantum critical trajectory ($g=g_c$) are shown in Fig.~\ref{f.xiTscaling-Ising}. There it is shown that both $\xi$ and $\xi_Q$ scale as $T^{-1/z}$ (with $z=1$) at low temperatures -- namely within the QC regime. In addition to the results for the quantum critical trajectory, we show the results for $g = 1.56 > g_c$, namely for a transverse field slightly above the quantum critical value. We observe that the $T$-dependence of the correlation length $\xi$ is extremely sensitive to the departure from the critical point, separating from the $g=g_c$ curve at a temperature ($T^* \approx 0.4 J$) higher than the gap (estimated to be $\Delta \approx 0.1 J$), to saturate to a finite value $\approx c/\Delta$ when $T \lesssim \Delta$. As a consequence, despite the small deviation from $g_c$, $\xi$ loses all traces of the scaling as $T^{-1}$ characteristic of the QC regime.
On the other hand the quantum coherence length for $g=1.56$ follows the same scaling as that at $g=g_c$ down to much lower temperatures ($T_Q^*\approx 0.06 J$), and this opens a finite temperature window over which the $T^{-1}$ scaling is manifest -- the same observation holds for the quantum variance, exhibiting the $T^{-\psi}$ scaling over the same temperature range (see main text). As already mentioned in the main text, this observation can be traced back to the fact that in the QC regime $\xi_Q \approx c_Q/T \ll \xi \approx c/T$, and hence, for $g>g_c$, the finiteness of $\xi_Q = \xi$ at $T=0$ leads to a deviation from the QC scaling at a temperature 
$T^*_Q \approx (c_Q/c) ~T^* \ll T^*$. This justifies the ability of quantum fluctuations to detect a QC regime of finite extent in temperature even when the ground state is quantum disordered. In the regime exhibiting an ordered ground state, the explanation of the sensitivity of quantum fluctuations to the QC regime is instead to be attributed to the weakness of the singularity exhibited by the QV at any thermal transition. As a consequence the Ginzburg region for quantum fluctuations is extremely small, and it affects minimally the temperature scaling of the QC regime above $T_c$.      

\begin{center} 
\begin{figure*} [ht!]
\includegraphics[width=0.8\textwidth]{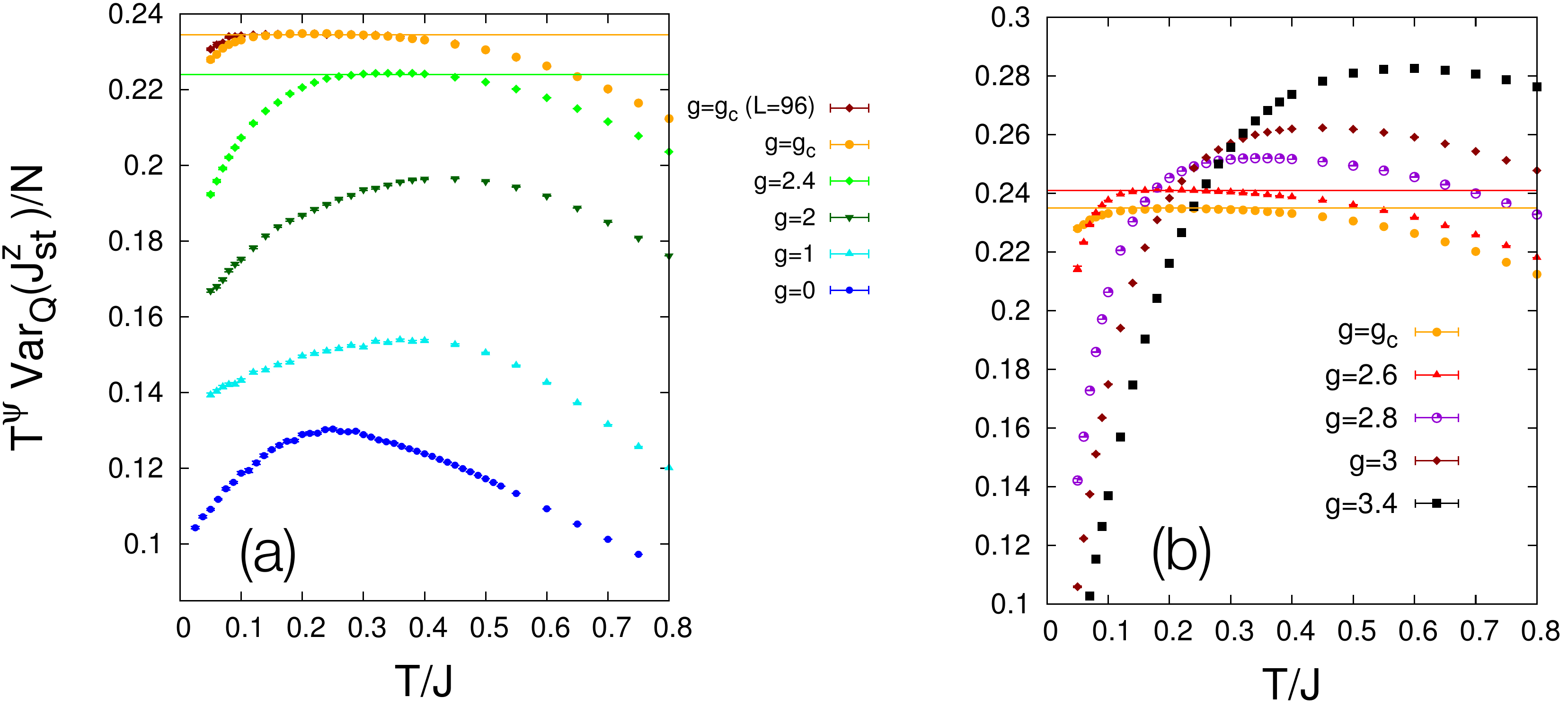}
\caption{Temperature scaling of the order-parameter quantum variance in the $S=1/2$ Heisenberg bilayer. Both panels show the temperature dependence of  $T^{\psi} {\rm Var}_Q(J^z_{\rm st})/N$ for a $N=L\times L \times 2$ lattice, below (a) and above (b) the quantum critical point. All data are for a $L=64$ lattice, except for the data in panel (a) at $g=g_c$ and $L=96$. The horizontal solid lines mark the plateaus exhibited by the curves in the quantum-critical regime. The low-$T$ downturn of the data for $g=g_c$ is a finite-size effect, as shown by the comparison between the data for $L=64$  and $L=96$ in (a).}  
\label{f.Tscaling-HeisBL}
\end{figure*} 
\end{center}  

\section{Temperature scaling of the quantum variance for the Heisenberg bilayer} 

In this section we discuss the temperature scaling of the order-parameter quantum variance for the Heisenberg bilayer and the specific signatures of the QC regime. Fig.~\ref{f.Tscaling-HeisBL} shows the function $T^{\psi} {\rm Var}_Q(J^z_{\rm st})/N$ for various values of the $g$ parameter below and above the QCP. In the vicinity of the QCP this function shows an extended plateau, manifesting the existence of a clear temperature interval characterized by the thermal QC scaling $\sim T^{\psi}$ -- the QC regime is then identified by this feature and the additional constraint that the prefactor to the $T^{-\psi}$ scaling must be close (within an $\epsilon$ tolerance) to that exhibited along the QC trajectory. 

 The plateau transforms into a broad maximum when leaving the QC regime -- namely the quantum variance exhibits a $T^{-\psi}$ temperature scaling only locally in $T$ (as its logarithmic derivative varies continuously with $T$). This is evident also in the single-layer limit $g=0$, reproducing the physics of the $S=1/2$ 2$d$ Heisenberg antiferromagnet. We conclude that the latter model does not possess evidence of an extended region of $T^{-\psi}$ scaling for the order-parameter quantum variance: namely quantum fluctuations do not detect any sign of a quantum-critical/renormalized-classical crossover in this system.

\section{Quantum Ising model in $d=1$ and $d=\infty$: quantum variance vs. quantum Fisher information}
 In this section we show the temperature scaling of the order parameter fluctuations, quantum variance and quantum Fisher information of the order parameter for the quantum Ising model in $d=1$ and $d=\infty$. Both cases are exactly solvable, which is an essential prerequisite for the calculation of the quantum Fisher information \cite{Haukeetal2016}. 
 
  We shall begin our discussion by the $d=\infty$ case, namely the fully connected Ising model, whose Hamiltonian reads
  \begin{equation}
  {\cal H}/{\cal J} = - \frac{\left (J^z \right)^2}{N} - g J^x
   \end{equation}
 Here we indicate with ${\cal J}$ the spin-spin coupling energy, not to be confused with the length of the collective spin ${\bm J} = \sum_i {\bm S}$. This model can be exactly solved by diagonalizing the Hamiltonian in each of the sectors defined by the conservation of ${\bm J}^2 = \hbar J (J+1)$, namely by diagonalizing the reduced Hamiltonian matrices ${\cal H}^{(J)}_{M,M'} = \langle JM | {\cal H} | JM' \rangle$. This analysis gives readily access to the eigenvalues $E_n$ and, when supplemented with the formula for the degeneracy of each $J$ sector \cite{arecchietal1972}, to all observables related to the collective spin. From this one can readily reconstruct the order parameter variance ${\rm Var}(J^z)$, the quantum variance of the order parameter \cite{FrerotR2016}
 \begin{equation}
 {\rm Var}_Q(J^z) = \sum_{nm} \left ( \frac{p_n + p_m}{2} - \frac{p_n - p_m}{\log(p_n/p_m)} \right ) \left | \langle n | J^z | m \rangle \right |^2 
 \end{equation}
 and its quantum Fisher information \cite{PezzeS2014}
 \begin{equation}
 {\rm QFI}(J^z) = \sum_{nm} 2  \frac{(p_n - p_m)^2}{p_n + p_m}  \left | \langle n | J^z | m \rangle \right |^2 
 \end{equation}
 where $p_n = \exp(-\beta E_n)/{\cal Z}$ with $\cal Z$ the partition function, and we have assumed that $\langle J^z \rangle = 0$ (valid on a finite-size system).  Introducing the two functions 
 \begin{eqnarray}
 G_{\rm QV}(x,y) & = &  \frac{x+y}{2} - \frac{x-y}{\log(x/y)} \nonumber \\
 G_{\rm QFI}(x,y) & = &  2 \frac{(x-y)^2}{x+y}  
 \end{eqnarray}
it is easy to show that  $G_{\rm QV}(x,y) \leq G_{\rm QFI}(x,y)/4 \leq 3 G_{\rm QV}(x,y)$ over the square $\{ x\in [0,1], y\in [0,1] \}$. This inequality chain is then inherited by the quantum variance and quantum Fisher information, namely 
\begin{equation}
{\rm Var}_Q(J^z) \leq {\rm QFI}(J^z)/4 \leq 3 {\rm Var}_Q(J^z)~.
\end{equation}
This implies \emph{e.g.} that  the quantum critical divergence of the quantum Fisher information at $g=g_c$ (for $T\to 0$ and for $N \to \infty$) is identical to that of the quantum variance. 

\begin{center} 
\begin{figure*} [ht!]
\includegraphics[width=0.89\textwidth]{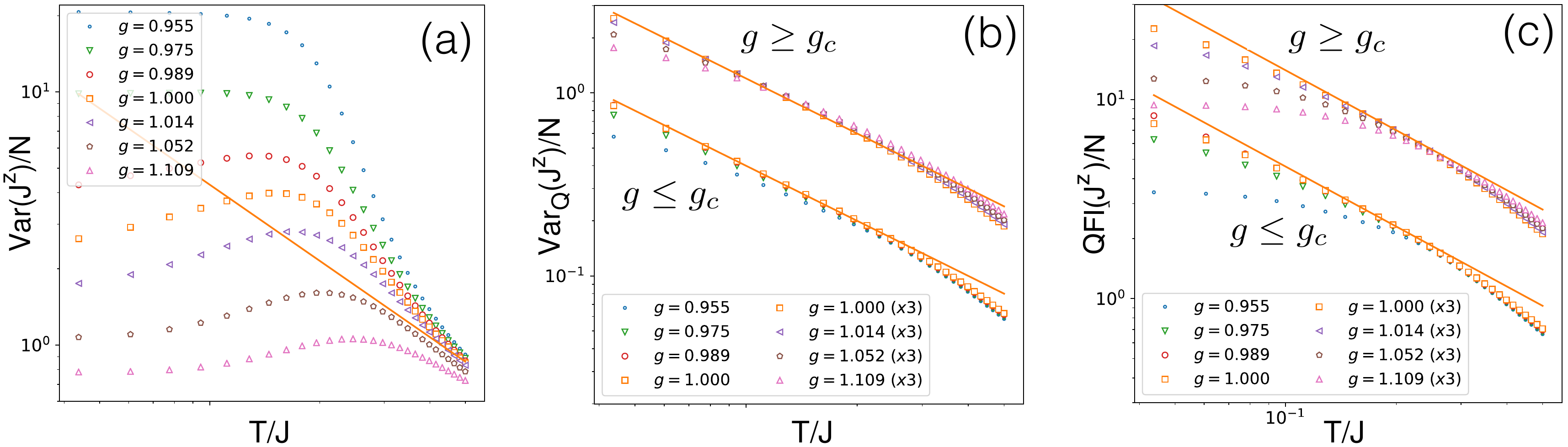}
\caption{Temperature scaling of the variance (a), quantum variance (b) and quantum Fisher information (c) of the transverse field Ising model in $d=\infty$. The data are obtained for a system size of $N=1000$.}  
\label{f.Tscaling-Isingdinf}
\end{figure*} 
\end{center}

 Fig.~\ref{f.Tscaling-Isingdinf} shows the temperature scaling of these three quantities around the quantum critical point of the model at $g=g_c=1$ for a system of size $N=1000$. Similarly to what seen in the main text for the case $d=2$, we observe that the $T$-dependence of the total variance strongly depends on $g$ around the quantum critical point. Moreover finite-size effects in this infinite-connectivity model are so strong that the expected power-law scaling along the quantum critical trajectory as $T^{-\psi}$ with $\psi =1$ (for mean-field exponents) is not observed. On the other hand, finite-size effects are much weaker on the quantum variance, which clearly exhibits the above power-law scaling at $g=g_c$; and, similarly to $d=2$, the same scaling is observed as well in an intermediate temperature range for $g\gtrsim g_c$ and $g \lesssim g_c$, marking the signature of the quantum critical fan. A rather similar behavior is also seen in the $T$-dependence of the quantum Fisher information,  although the latter appears to be more affected by finite-size effects; and it seems to be more sensitive to the deviation from the quantum critical point, suggesting a smaller quantum critical fan. We notice that our calculation of the quantum Fisher information differs from that of Ref.~\cite{Haukeetal2016} in that the latter was limited to the Hamiltonian sector $J=N/2$, making the model identical to that of the bosonic Josephson junction \cite{Pezzeetal2016}. 
 
\begin{center} 
\begin{figure*} [ht!]
\includegraphics[width=0.9\textwidth]{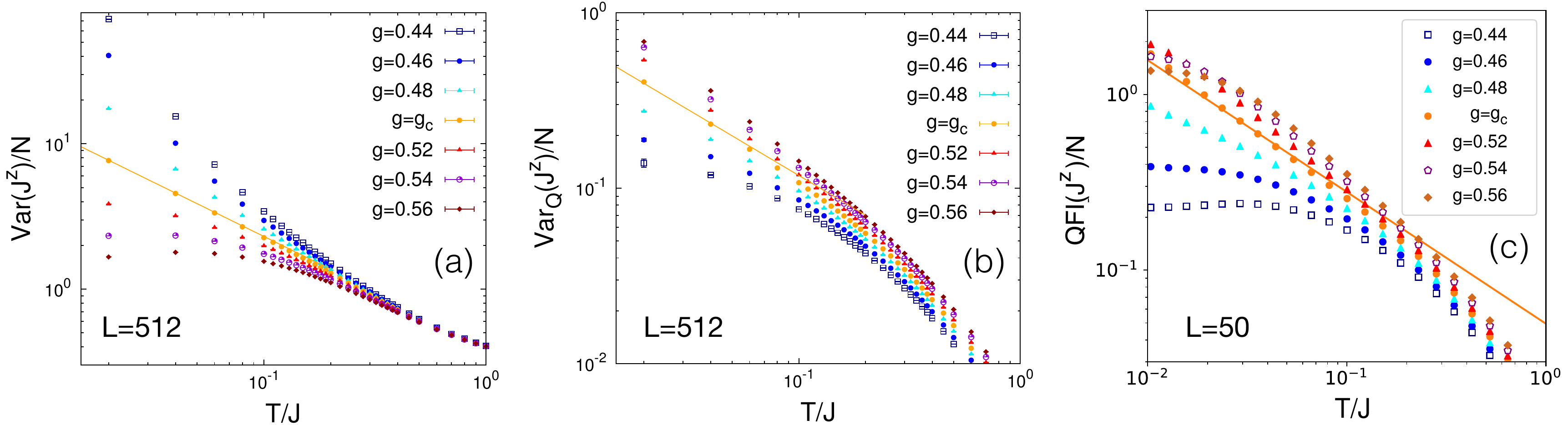}
\caption{Temperature scaling of the variance (a), quantum variance (b) and quantum Fisher information (c) of the transverse field Ising model in $d=1$. The data in (a) and (b) were obtained via quantum Monte Carlo for a chain of length $L=512$ with periodic boundary conditions. The data in (c) were calculated for a chain of length $L=50$ with open boundary conditions.}  
\label{f.Tscaling-Ising1d}
\end{figure*} 
\end{center}

  We conclude by discussing the case $d=1$, which can be  solved exactly via a Jordan-Wigner transformation to free fermions, and which exhibits a quantum critical point for $g=g_c = 1/2$.  The $1d$ case is special in that the system no-longer exhibits a finite-$T$ transition on the ordered side $g<g_c$.  The exact solution gives access to the dynamical susceptibility $\chi''$ for the order parameter, which allows then to reconstruct the quantum Fisher information via the formula of Ref.~\cite{Haukeetal2016}. At finite temperature the exact solution is only viable for open boundary conditions, and it remains rather laborious (implying the calculation of Pfaffians of large matrices) in order to reconstruct correlation functions. It is therefore useful to make use of the numerical solution via quantum Monte Carlo for the quantities which are accessible to this method, namely the total and quantum variance. 
  
  The temperature dependence of the latter two quantities are shown in Fig.~\ref{f.Tscaling-Ising1d}(a-b) for a chain of length $L=512$ with periodic boundary conditions. There we observe that a quantum-critical $T^{-\psi}$ scaling (with $\psi = 3/4$) is clearly exhibited by the variance along the quantum critical trajectory for $T/J \lesssim 0.1$. Yet the total variance shows also a strong sensitivity to a deviation from the quantum critical point, and, over the above temperature range, the power-law behavior is quickly lost when $g$ deviates from $g_c$. The quantum variance shows less sensitivity to a deviation from $g_c$, but, contrary to $d>1$, its power-law scaling along the quantum critical trajectory shows up on a smaller temperature range than for the total variance; as a consequence, in the data presented in Fig.~\ref{f.Tscaling-Ising1d} the quantum critical behavior of the quantum variance is not observed away from $g_c$. We also remark that the $d=1$ case is rather special for the quantum Ising model in that, over the temperature range shown in the figure, the quantum variance is \emph{not} maximal along the quantum critical trajectory (compare Fig.~\ref{f.Tscaling-Isingdinf}(b) of the present Supplementary Material, and Fig.~2(b) of the main text), but rather for $g>g_c$. 
  
 Interestingly all the remarks made for the quantum variance also hold for the quantum Fisher information, shown in Fig.~\ref{f.Tscaling-Ising1d}(c) for a open-boundary chain of length $L=50$. 
 We conclude therefore that, in the specific case of the 1d quantum Ising model, the temperature scaling of quantum correlations does not exhibit a quantum critical regime which is significantly broader in $g$-$T$ space than that shown by conventional correlations, and which for both forms of correlations appears to be tightly confined around the quantum critical trajectory. The upper bound to the $T^{-\psi}$ quantum critical scaling of the total variance along such a trajectory ($T/J \approx 0.1$) is consistent with the upper bound to the QC regime exhibited by the temperature scaling of the free energy \cite{KoppC2005}.

\bibliography{QCfan.bib}

\end{document}